\documentclass[aps,  prd,twocolumn,  superscriptaddress]{revtex4-2}

\usepackage[english]{babel}
\usepackage{amsmath}
\usepackage{graphicx}
\usepackage[colorlinks=true, allcolors=blue]{hyperref}
\usepackage{dcolumn}
\usepackage{lineno}
\usepackage{multirow}
\usepackage{subfigure}
\usepackage{soul}
\usepackage{comment}
\usepackage{float}
\usepackage{ulem}
\begin{document}
\title{Neutrino type identification for atmospheric neutrinos in a large homogeneous liquid scintillation detector}

%\author{Authors}
\newcommand{\SDU}{Shandong University, Jinan, China, and Key Laboratory of Particle Physics and Particle Irradiation of Ministry of Education, Shandong University, Qingdao 266237, China}
\newcommand{\IHEP}{Institute of High Energy Physics, Chinese Academy of Sciences, Beijing 100049, China}
\newcommand{\UCAS}{School of Physical Sciences, University of Chinese Academy of Science, Beijing 100049, China}

\author{Jiaxi Liu}
\altaffiliation{Jiaxi Liu and Fanrui Zeng contributed equally to this work.}
\affiliation{\IHEP}
\affiliation{\UCAS}
\author{Fanrui Zeng}
\altaffiliation{Jiaxi Liu and Fanrui Zeng contributed equally to this work.}
\affiliation{\SDU}
\author{Hongyue Duyang}
\email{Corresponding author: duyang@sdu.edu.cn}
\affiliation{\SDU}
\author{Wanlei Guo}
\affiliation{\IHEP}
\author{Xinhai He}
\affiliation{\IHEP}
%\affiliation{\UCAS}
\author{Teng Li}
\email{Corresponding author: tengli@sdu.edu.cn}
\affiliation{\SDU}
\author{Zhen Liu}
\email{Corresponding author: liuzhen@ihep.ac.cn}
\affiliation{\IHEP}
\author{Wuming Luo}
\email{Corresponding author: luowm@ihep.ac.cn}
\affiliation{\IHEP}
\author{Wing Yan Ma}
\email{Corresponding author: wingyanma@sdu.edu.cn}
\affiliation{\SDU}
\author{Xiaohan Tan}
\affiliation{\SDU}
\author{Liangjian Wen}
\affiliation{\IHEP}
\author{Zekun Yang}
\affiliation{\SDU}
\author{Yongpeng Zhang}
\email{Corresponding author: ypzhang1991@ihep.ac.cn}
\affiliation{\IHEP}

\begin{abstract}

Atmospheric neutrino oscillations are important to the study of neutrino properties, including the neutrino mass ordering problem. 
A good capability to identify neutrinos' flavor and neutrinos against antineutrinos is crucial in such measurements. 
In this paper, we present a machine-learning-based approach for identifying atmospheric neutrino events in a large homogeneous liquid scintillator detector.  
This method identifies features of PMT waveforms that reflect event topologies and uses them as input to machine learning models. 
In addition, neutron-capture information is utilized to achieve neutrino versus antineutrino discrimination. 
Preliminary performances based on Monte Carlo simulations are presented, which demonstrate such a detector's potential in future measurements of atmospheric neutrinos such as the one planned for the JUNO experiment. 
\end{abstract}

\maketitle

\section{Introduction}

Atmospheric neutrinos constitute an important natural neutrino source for investigating fundamental neutrino properties.  
Produced by cosmic rays interacting with the Earth's atmosphere, atmospheric neutrinos span wide energy spectra and oscillation baseline lengths, making them well-suited for studying neutrino oscillations.  
The most notable achievement made by atmospheric neutrino measurements is the discovery of neutrino oscillations by the Kamiokande~\cite{Kamiokande-II:1988sxn} and Super-Kamiokande experiments~\cite{Super-Kamiokande:1998kpq}. 
One of the biggest questions in neutrino physics that future atmospheric neutrino measurements could solve is the sign of $\Delta m^2_{31}$ and $\Delta m^2_{32}$, commonly known as the neutrino mass ordering (NMO) (or neutrino mass hierarchy) problem. 
Matter effects, due to charge-current interactions between neutrinos and electrons when atmospheric neutrinos pass through the earth, could help experiments solve the problem by amplifying the differences in oscillation probabilities between NO and IO~\cite{Wolfenstein:1977ue,MSW:1986wj}.
The interpretations of such measurements, however, are also affected by uncertainties such as the unknown value of the CP-violating phase ($\delta_{\mathrm{CP}}$) and the $\theta_{23}$ octant ambiguity.
The next-generation Cherenkov neutrino experiments, such as Hyper-Kamiokande~\cite{Hyper-Kamiokande:2011ts,Hyper-Kamiokande:2018ofw}, KM3Net/ORCA~\cite{KM3Net:2016zxf, KM3NeT:2021ozk} and IceCube-Upgrade~\cite{ICECUBE_Upgrade:2019aao}, capable of measuring atmospheric neutrino oscillations and solve the NMO problem with gigantic water detectors, are currently under construction. 

A different method that can solve the NMO problem is the measurement of medium baseline ($\sim$50 km) reactor antineutrino oscillation in vacuum or near-vacuum conditions. 
This method has the advantage that it does not depend on $\delta_{\mathrm{CP}}$ and the $\theta_{23}$ octant. 
The only experiment on the horizon with this method is the Jiangmen Underground Neutrino Observatory (JUNO), currently under construction in China.
It is designed to determine NMO with a 20\,kton homogeneous liquid scintillator (LS) detector by measuring reactor electron antineutrino ($\bar{\nu}_e$) oscillations \cite{JUNOYB,JUNO_phydet}. 
It is also worth noting that the two methods are complementary, and the synergy between them can greatly enhance the total sensitivity in a global fit \cite{IceCube-Gen2:2019fet,Cabrera:2020ksc,KM3NeT:2021rkn}. 
Therefore, if atmospheric (anti)neutrino oscillations can also be measured in a large homogeneous LS detector like JUNO itself, 
its NMO sensitivity can be maximized by a joint analysis of both reactor and atmospheric neutrino oscillations in the same detector. 

Atmospheric neutrino oscillation measurements in an LS detector, however, are very challenging. 
While such detectors has been utilized by experiments such as KamLAND \cite{Kamland}, Borexino \cite{Borexino,BOREXINO:2018ohr}, Daya Bay \cite{DayaBay:2012fng, DayaBay:2022orm}, RENO \cite{RENO:2012mkc,RENO:2018dro}, and Double Chooz \cite{DoubleChooz:2011ymz,DoubleChooz:2019qbj} for reactor and solar neutrino measurements, they have never been used for atmospheric neutrinos. 
This is mostly due to two difficulties. 
First, LS detectors are traditionally believed to lack directional measurement capability, which is mandatory for atmospheric neutrino oscillation measurements to determine the neutrino oscillation baseline length.  
Second, the atmospheric neutrino flux is a mixture of muon (anti)neutrino ($\nu_\mu$/$\bar{\nu}_\mu$) and electron (anti)neutrinos ($\nu_e$/$\bar{\nu}_e$) and hence the neutrino flavor needs to be identified in the detector to determine the oscillation probability. 
Moreover, the ability to identify neutrinos against antineutrinos is important since matter effects distort their oscillation probability in opposite ways.
While experiments lacking such capability still get some sensitivity to NMO from small differences in fluxes and cross sections for neutrinos and antineutrinos, the discriminating power between the neutrinos and antineutrinos certainly helps to disentangle NMO and CP violation phase and increase the sensitivity.  
However, such flavor identification and neutrino/antineutrino discrimination have never been achieved in LS detectors before. 

Neutrino flavors are reflected in the flavors of outgoing charged leptons in neutrino-nucleus charged current (CC) interactions. 
Therefore, a detector's capability of neutrino flavor identification largely relies on the identification of charged leptons, \textit{i.e.}, $e^{\mp}$ in $\nu_e$/$\bar{\nu}_e$-CC, and $\mu^{\mp}$ in $\nu_\mu$/$\bar{\nu}_\mu$-CC interactions. 
In a water Cherenkov detector, this task is achieved by identifying the sharpness of the Cherenkov ring.
But this approach is not feasible in an LS detector, where Cherenkov light is typically about two orders of magnitude weaker than scintillation light.
Neutrino versus antineutrino identification is even more challenging. 
The most prominent difference between neutrino and antineutrino interactions is the charge of outgoing leptons in CC interactions. 
However, it is impractical to magnetize a modern neutrino detector with a kilotons-scale mass to identify the particle charge. 
Other handles may help in this task including the differences in event kinematics and neutron multiplicity in the (anti)neutrino interactions. 
One recent example in a water Cherenkov detector is the atmospheric neutrino oscillation measurement by Super-Kamiokande with neutron-tagging~\cite{Super-Kamiokande:2023ahc}.
Super-Kamiokande with gadolinium loading is expected to improve further the neutrino/antineutrino discrimination with better neutron-tagging efficiency~\cite{Super-Kamiokande:2021the,Super-Kamiokande:2024kcb}.  
Besides, event inelasticity, defined as the fraction of energy transferred from the incoming neutrino to the hadrons and exhibits different shapes between neutrinos and antineutrinos, is also proposed to be used in IceCube-upgrade and KM3NET/ORCA to separate neutrinos/antineutrinos and improve the oscillation sensitivity to NMO~\cite{Ribordy:2013xea,Olavarrieta:2024eaq}. 
LS detectors, on the other hand, are known to have excellent neutron-tagging efficiency, but event kinematics such as inelasticity measurements are very difficult with traditional reconstruction methods.

Recently, we reported a novel method of directionality measurement in an LS detector that solves the difficulty of directionality measurement in utilizing an LS detector for atmospheric neutrino oscillation measurements \cite{Yang:2023rbg}. 
This method extracts features relevant to the event topology in the detector from PMT waveforms and uses them as input to machine learning (ML) models trained to find the event directionality. 
In this paper, we further explore the possibility of flavor identification in an LS detector such as JUNO for atmospheric neutrinos with a methodology based on a similar ML principle. 
A statistical neutrinos versus antineutrinos identification is also demonstrated with additional help from neutron capture information. 

This paper is organized as follows. 
Section \ref{sec:det_sim} describes the detector and simulation used in the work. 
Section \ref{sec:3label} presents the methodology of 3-label identification, i.e., $\nu_\mu$/$\bar{\nu}_\mu$-CC versus $\nu_e$/$\bar{\nu}_e$-CC versus neutral current (NC) identification, 
and section \ref{sec:nu_nubar} presents neutrino vs anti-neutrino identification.  
The identification performance is presented in section \ref{sec:performance}, and more details including the performance's dependence on event generators, score cut optimization, and an alternative classification strategy are further discussed in section \ref{sec:disscusion}.
Finally section \ref{sec:sum} summarizes the study. 

\section{Detector and Simulation \label{sec:det_sim}}

\begin{figure}[htbp]
\centering
\includegraphics[width=0.49\textwidth]{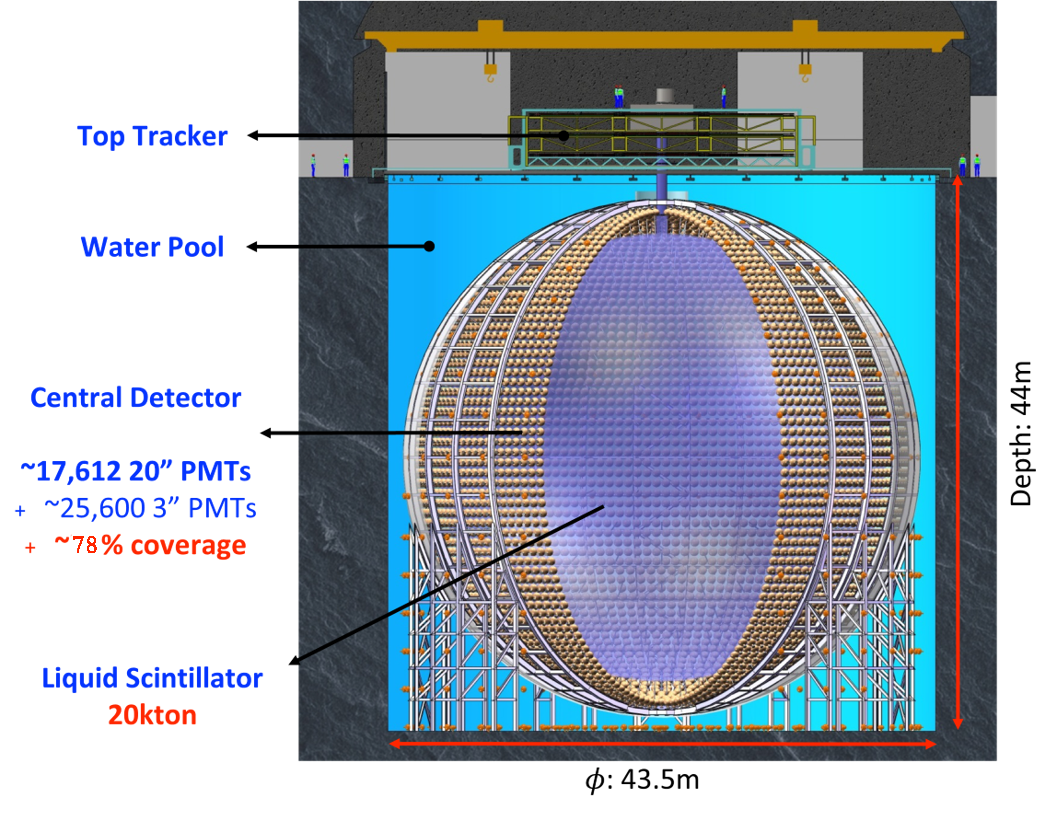}
\caption{\label{fig:detector} Drawing of the JUNO detector design. }
\end{figure}

\begin{figure}[htbp]
\centering
\includegraphics[width=.45\textwidth]{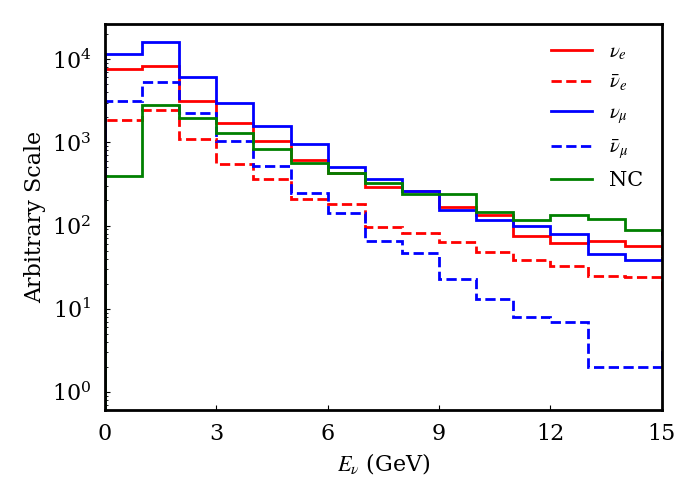}
\caption{The neutrino energy distribution of the Honda-flux~\cite{Honda:2015fha} sample used for performance evaluation in this work. The sample is divided into five classes: $\nu_\mu$/$\bar{\nu}_\mu$/$\nu_e$/$\bar{\nu}_e$-CC and NC, represented by different colors.  \label{fig:honda_samples}}
\end{figure}

The JUNO central detector (CD) design and simulation~\cite{JUNO:2023ete,Lin:2022htc} are used to demonstrate the method.
A schematic view of the detector is shown in Fig.~\ref{fig:detector}.
It contains 20\,kton LS in an acrylic sphere with a radius of 17.7 m.  
17,612 20-inch PMTs~\cite{Coppi:2023nlv} are installed facing inward to collect light produced by charged particles with a total PMT coverage of 75\%. 
In addition, 25,600 3-inch PMTs~\cite{JUNO_3inchPMT} provide additional 3\% coverage. 
For simplicity, only 20-inch PMTs are used in this work. 
The LS consists of linear alkylbenzene as the detection medium, 2.5\,g/L 2,5-diphenyloxazole (PPO) as the fluor and 3 mg/L p-bis- (o-methylstyryl)-benzene (bis-MSB) as the wavelength shifter~\cite{JUNO_LS}.  
More details about the JUNO detector can be found in Ref.~\cite{JUNOYB}.

Neutrino interactions in the detector are simulated mainly by the GENIE event generator (v3.0.6)~\cite{Andreopoulos:2015wxa,Andreopoulos:2021prd}, and the NuWro event generator~\cite{Golan:2012rfa} is used as an independent check. 
The detector response is simulated based on Geant4~\cite{GEANT4:2002zbu,Allison:2006ve}. 
Various electronic effects, including PMT dark noise, time transition spread (TTS), PMT charge smearing, electronics baseline fluctuation and PMT response for single photoelectron, are also implemented. 
The detailed PMT simulation parameters are the same as in Ref.~\cite{Yang:2023rbg}.

Three independent neutrino interaction samples are produced to train and validate the ML models, each including five categories of events: $\nu_\mu$-CC, $\bar{\nu}_\mu$-CC, $\nu_e$-CC, $\bar{\nu}_e$-CC and NC.
The first sample is simulated with GENIE and has roughly flat visible energy distributions and similar statistics (about 70\,k) for each category to avoid introducing any energy dependence in the model training (referred to as the flat sample). 
The second is simulated with GENIE and a more realistic flux taken from the calculation of Honda \textit{et al.}~\cite{Honda:2015fha} for the JUNO site without the overburden mountain (referred to as the Honda-flux sample). 
This sample is used for performance evaluation with total statistics of about 95\,k. 
The neutrino energy spectrum of the Honda-flux sample is shown in Fig. \ref{fig:honda_samples}.
The last one is simulated with NuWro with similar statistics to the Honda-flux sample to test the robustness of the performance with a different generator (referred to as the NuWro sample). For each sample, events with visible energy between 0.5 and 15\,GeV are selected for this study, which covers the most sensitive energy region of atmospheric neutrinos to NMO~\cite{JUNOYB}. 

\begin{figure*}[htbp]
\centering
\subfigure[muon]{\hspace{-0.5cm}
\includegraphics[width=.45\textwidth]{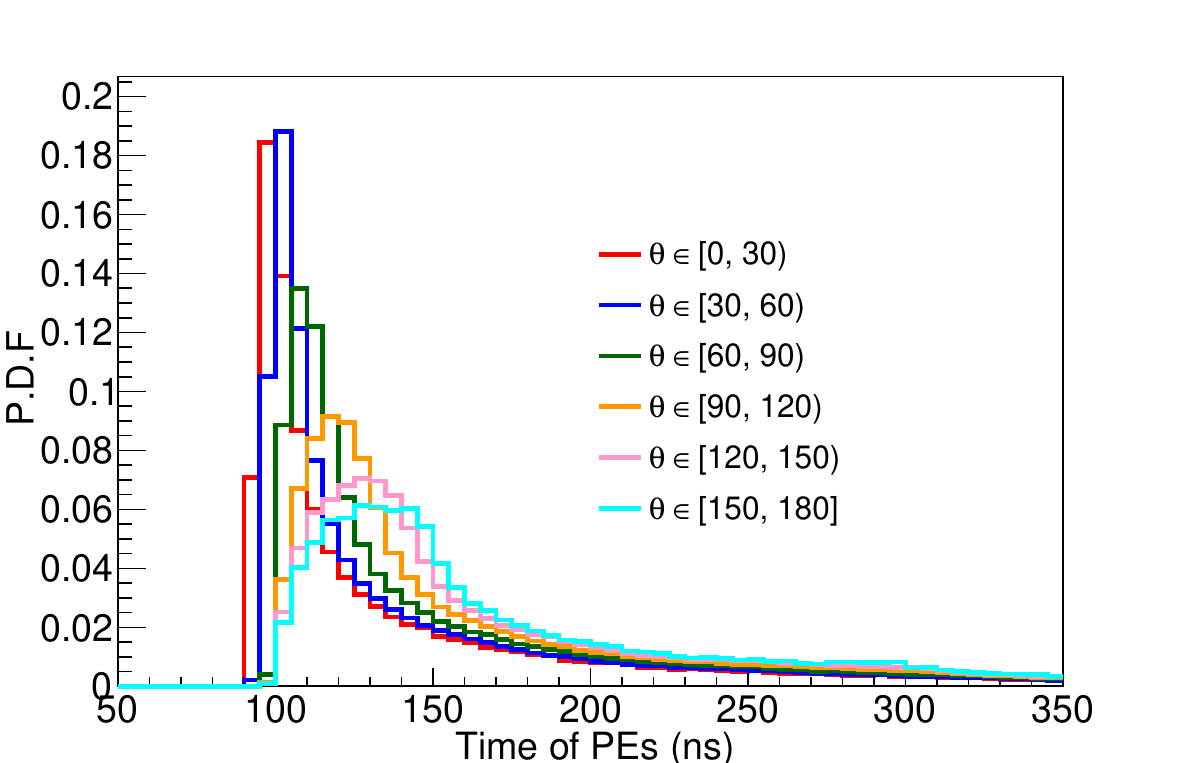}
}
\qquad
\subfigure[electron]{\hspace{-0.5cm}
\includegraphics[width=.45\textwidth]{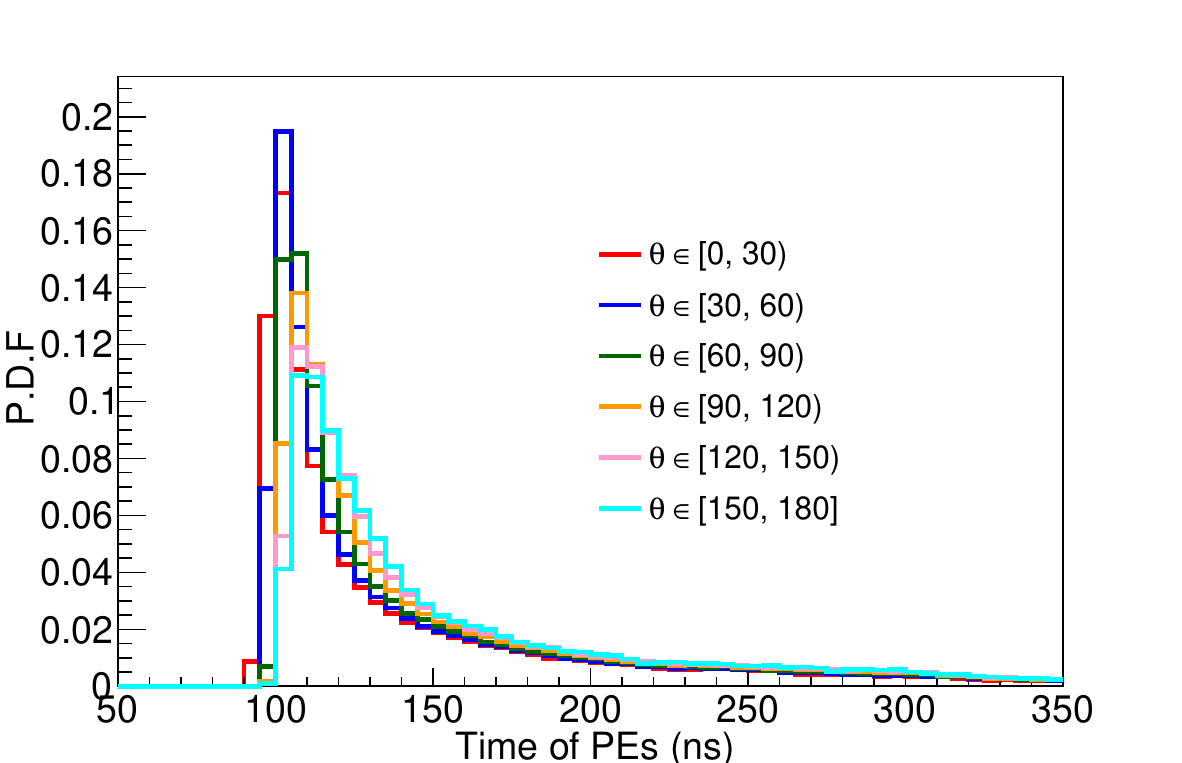}
}
\caption{The normalized time distributions of PEs for PMTs at different angles ($\theta$, in degree) to (a) a muon and  (b) an electron with the same kinetic energy (1\,GeV), vertex position (at the center of the detector) and initial direction (along the positive x-axis). The PMT angle $\theta$ is the intermediate angle between the particle direction and the line connecting the middle point of the particle track and the PMT. Distinct shapes can be observed for the two types of particles. \label{fig:npe_t}}

\end{figure*}
\begin{figure*}[htbp]
\centering
\includegraphics[width=0.45\textwidth]{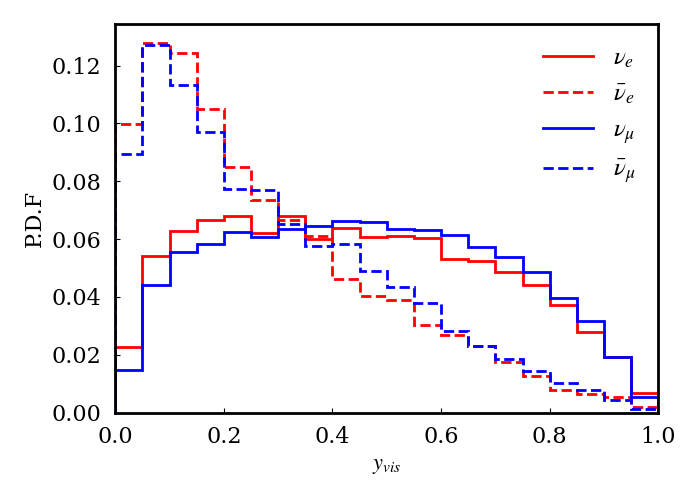}
\includegraphics[width=0.45\textwidth]{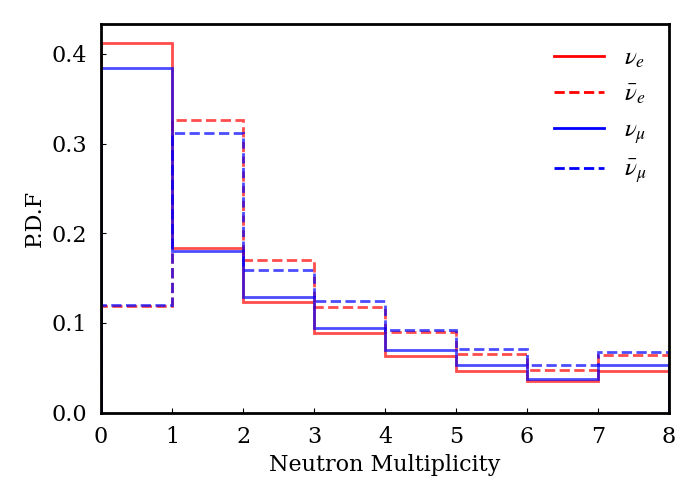}
\caption{\label{fig:y_nmultiplicity} $y_{vis}$ (left) and neutron multiplicity (right) distributions for $\nu_\mu$/$\bar{\nu}_\mu$/$\nu_e$/$\bar{\nu}_e$-CC interactions in the Honda-flux sample. Full detector simulation is included. The distributions are normalized for a shape-only comparison. }
\end{figure*}

\begin{figure*}[htbp]
\centering
\includegraphics[width=0.3\textwidth]{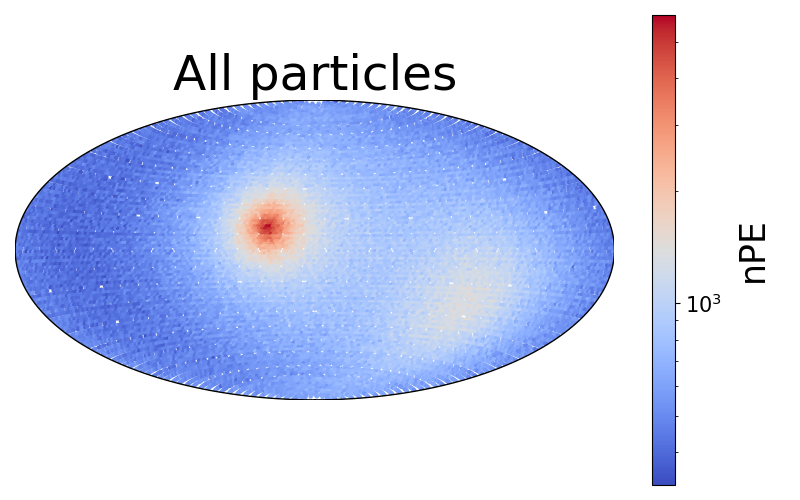}
\includegraphics[width=0.3\textwidth]{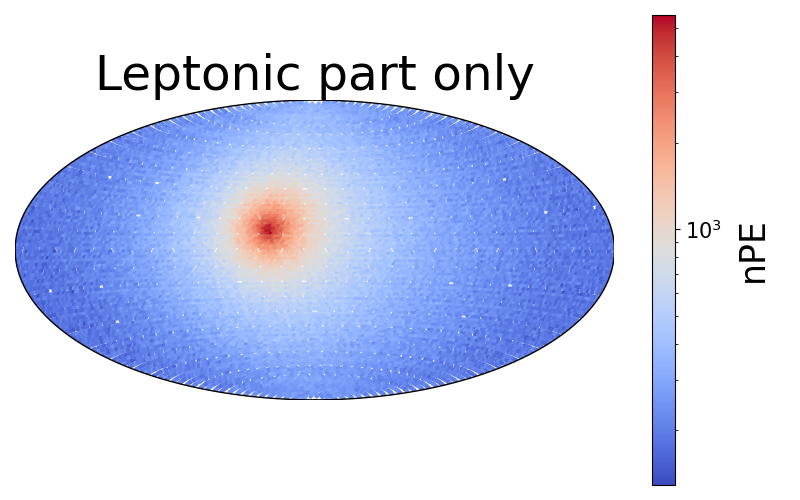}
\includegraphics[width=0.3\textwidth]{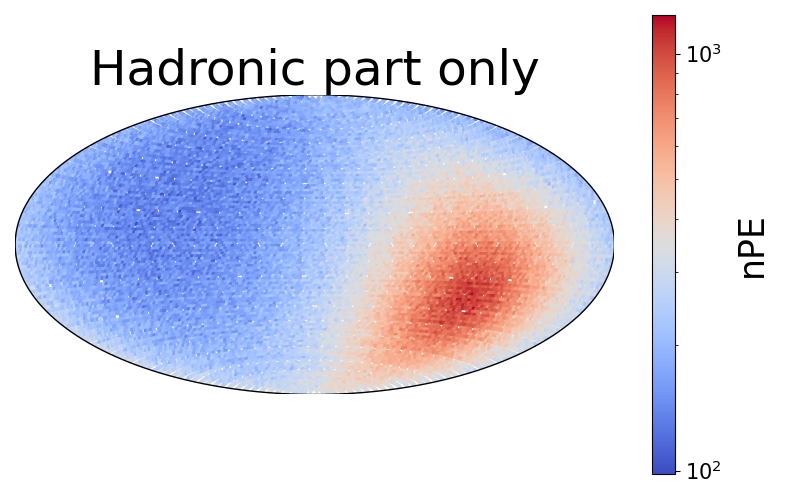}
\caption{Event display pictures of a simulated $\nu_e$-CC interaction showing the PMT charges (nPE) from all particles (left), leptonic part only (middle), and hadronic part only (right). The PMT charge values are represented by colors and are projected onto a planer surface following the Mollweide projection method.   \label{fig:evd} } 
\end{figure*}

\begin{figure*}[htbp]
\centering
\includegraphics[width=0.9\textwidth]{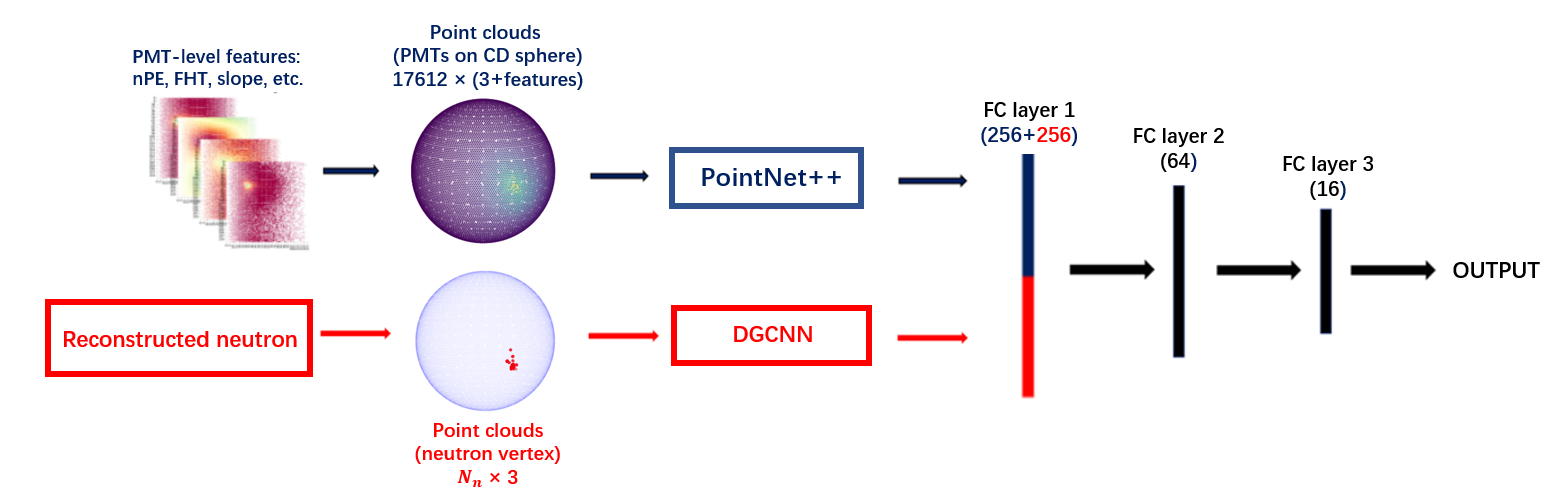}
\caption{\label{fig:model1} Schematic diagram of the model architecture used by $\nu/\bar{\nu}$ identification strategy 1. The model consumes two types of inputs: PMT features from the prompt trigger processed by PointNet++, and neutron-capture vertex cloud processed by DGCNN. The outputs of PointNet++ and DGCNN are input to fully connected (FC) layers to predict the final scores representing the probability of the event being caused by neutrinos or antineutrinos. }
\end{figure*}

\begin{figure*}[htbp]
\centering
\includegraphics[width=0.9\textwidth]{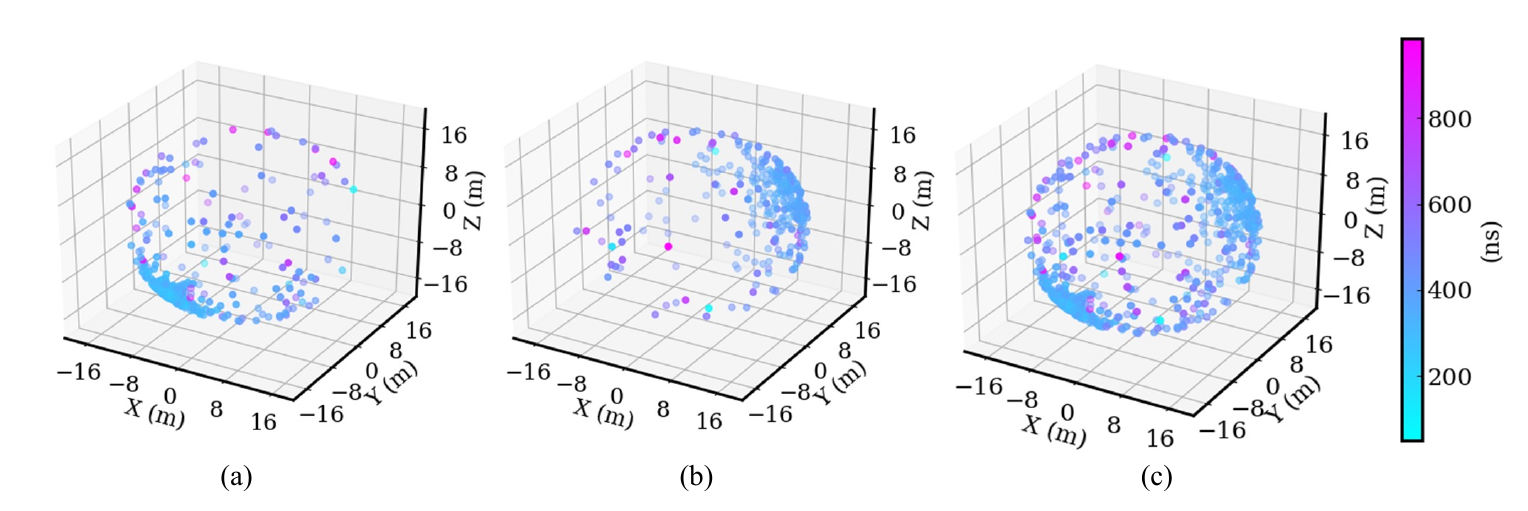}
\caption{\label{fig:strategy2_merge} Illustration of the trigger-merging technique used by strategy 2. (a) and (b) show two delayed triggers selected as neutron-capture candidate triggers, and (c) shows the new merged one. Fired PMTs are shown as dots on the figures, with color representing FHT values of the hits.}
\end{figure*}

\section{Methodology} \label{sec:strategy}
Neutrino-nucleus interactions in an LS detector are complicated in the few-GeV energy region.
In a typical $\nu_\mu$/$\bar{\nu}_\mu$-CC event, the $\mu^\mp$ from the primary interaction deposits its energy quickly and leaves a long track in LS. 
Various hadrons such as $n$, $p$, $\pi^0$ and $\pi^{\pm}$ can also be produced through primary and secondary interactions. 
Most charged hadrons and $\pi^0$s produce showers while rapidly depositing their energies in LS, with the exception that $\pi^{\pm}$s behaves similarly to $\mu^{\pm}$s.  
For a $\nu_e$/$\bar{\nu}_e$-CC event, the event topology in the LS is similar to that of $\nu_\mu$/$\bar{\nu}_\mu$-CC except that $e^\mp$ from the primary interaction also produces a shower. 
The primary charged lepton and hadrons together trigger the detector promptly (the prompt trigger). 
For an NC event, the hadron part is similar to a CC event, yet the outgoing neutrino is invisible.

Tracks and showers propagate through LS and produce scintillation light. 
Scintillation light seen by a PMT is the superposition of light from many points along the track or shower.
The time evolution of scintillation light seen by PMTs is therefore determined by the event topology in the detector. 
Fig. \ref{fig:npe_t} shows the PE time distributions ($n_{PE}(t)$) of PMTs at different angles with respect to an electron and a muon simulated at the detector center with the same kinetic energy (1\,GeV) and direction. 
It is observed that the two particles exhibit distinct $n_{PE}(t)$ shapes. 
This is mostly because electron showers are generally much shorter than muon tracks at the same energy in the LS. 
The differences in the $n_{PE}(t)$ distribution are then reflected in the PMT waveforms. 
Therefore, it is possible to use the information from PMT waveforms of the prompt trigger to identify different event topologies and distinguish the $\nu_\mu$/$\bar{\nu}_\mu$-CC, $\nu_e$/$\bar{\nu}_e$-CC, and NC atmospheric neutrino events.

While it is impossible to discriminate neutrino and antineutrino interactions event by event without identifying the charge of the outgoing lepton, the task can be done statistically with help from several handles.
The first one is their differences in event kinematics.
In CC interactions, neutrinos tend to transfer more energy to hadrons than anti-neutrinos with the same energy due to the V-A structure of $\nu$/$\bar{\nu}$ interactions. % ~\cite{xx}. 
This leads to a larger visible hadronic energy fraction for neutrino interactions (Fig. \ref{fig:y_nmultiplicity} left), defined as $y_{vis} = E_{had,vis}/E_{vis}$, where $E_{had,vis}$ is the visible energy caused by hadrons and $E_{vis}$ is the total visible energy of the event.
One advantage of LS detectors over water Cherenkov detectors is that the hadrons produced from atmospheric neutrinos are also mostly visible. 
To help illustrate this point, Fig.~\ref{fig:evd} shows an example of a simulated neutrino interaction with explicit separation between the PMT charges induced by the final-state lepton and hadrons. 
Since both the charged leptons and most hadrons in the final state of CC events deposit their energy within the prompt trigger, the $y_{vis}$ information is also reflected in the event topology and finally also in the trigger's PMT waveforms. 

The second handle is the neutron information. 
Neutrons can be produced both in the primary neutrino-nucleus interactions and also in secondary interactions.
Antineutrino interactions generally tend to produce more primary neutrons %since charge conservation in antineutrino (neutrino) charged-current processes converts protons (neutrons) into neutrons (protons), accompanied by positron (electron) production.
since charge conservation in an antineutrino CC interaction requires a negative charge being transferred from the lepton to the hadrons and is more likely to convert a proton into a neutron. In contrast, a positive charge is transferred in a neutrino CC interaction and is more likely to convert a neutron into a proton.
%due to charge conservation. 
The number and spacial distribution of secondary neutrons are also related to the types and kinematics of the primary particles. 
Neutron multiplicities (including both primary and secondary neutrons) for different neutrino CC interactions are shown in Fig.~\ref{fig:y_nmultiplicity} (right). 
The neutrons produced dissipate their kinetic energy and are eventually captured mostly by hydrogen nuclei with a time constant of about 200\,$\mu$s.
The average distance from the primary neutron capture position to the neutrino interaction vertex is about 0.5\,m in LS according to simulation, and the capture positions for secondary neutrons are much more dispersed. 
2.2\,MeV $\gamma$s are then released and create delayed triggers.  
Another advantage of LS detectors such as JUNO is the excellent efficiency ($>$90\%) in tagging such neutron captures compared to Water Cherenkov detectors. 
The information from delayed triggers caused by neutron captures therefore also helps to distinguish neutrinos from antineutrinos. 
More details on how to utilize the delayed trigger information will be discussed in Sec.~\ref{sec:nu_nubar}. 

Based on the discussion above, the identification task of atmospheric neutrino interactions is broken down into two steps:  
firstly to identify the $\mu$-like ($\nu_\mu$/$\bar{\nu}_\mu$-CC), e-like ($\nu_e$/$\bar{\nu}_e$-CC), and NC-like events against each other 
(referred to as the 3-label identification), 
and secondly to discriminate $\nu$-like from $\bar{\nu}$-like events (referred to as the 2-label identification).
The details of each step are described in the following subsections.
Sec.~\ref{sec:disscusion} also briefly discusses other possible strategies.

\subsection{3-label flavor identification}
\label{sec:3label}
In this step, atmospheric neutrino events are classified as 
$\mu$-like, $e$-like, or NC-like
using information from the 20-inch PMTs' waveforms in the prompt triggers. 
%A machine-learning-based method similar to that in Ref.~\cite{Yang:2023rbg} is used. 
Features are extracted from PMT waveforms with a sampling rate of 1~GHz in a 1000\,ns trigger window as inputs to ML models. 
These features include the hit time of the earliest photon (first hit time, FHT), the total charge, the slope of the waveform in the first 4\,ns, the charge ratio in the first 4\,ns to the total, the time and amplitude of the waveform's maximum bin, and other features such as median time and four moments (mean, std, skewness, kurtosis) of the waveform  which describe the reconstructed waveforms in a more subtle way.
A deconvolution and noise-reduction algorithm~\cite{deconv, HUANG201848} is applied to the waveform before the feature extraction to improve the feature quality. 

The PMT features from all 20-inch PMTs of the JUNO CD form spherical image-like data.
Two different ML models are used to deal with this kind of data: the Deepsphere~\cite{DeepSphere} model which processes spherical data directly and the PointNet++~\cite{Pointnet2} model which inputs data as a 3D point cloud. 
The model architectures remain mostly the same except that 
the activation functions of the last layer are replaced with the softmax function to output three scores representing the probability of the event being $\mu$-like, $e$-like,or NC-like respectively. 
The sum of these three scores is one and an event is classified to the category with the largest score by default.

\subsection{2-label $\nu$/$\bar{\nu}$ identification }
\label{sec:nu_nubar}

The identification of $\nu$/$\bar{\nu}$ interactions relies on their differences in event kinematics, neutron multiplicity, and the spatial distribution of neutron-capture vertices. 
The event kinematics information contained in the PMT waveforms of the prompt trigger is also reflected in the waveform features described in the previous subsection, 
including the $y_{vis}$ information discussed in section \ref{sec:strategy}. 
The neutron information, on the other hand, is reflected in the delayed triggers which are selected with energy between 2 and 2.7\,MeV, and the delay time between 10\,us and 1\,ms. 
Therefore, the information from the prompt and delayed triggers needs to be combined for the task. 
Two different strategies are developed and described here. 
For both strategies, the final output is a 2-label classifier, with 2 scores representing the probability of an event being $\nu$-like or $\bar{\nu}$-like respectively.
The models are then trained separately for the $\nu_\mu$ vs $\bar{\nu}_\mu$ and $\nu_e$ vs $\bar{\nu}_e$ identification. 

\subsubsection{Strategy 1}
In this strategy, neutron-capture vertices are first reconstructed using the method from Ref.~\cite{Huang:2022zum}. 
The reconstructed vertices from multiple selected delayed triggers form a new 3D point cloud that is input to the ML model together with the one formed by PMT waveform features from the prompt trigger. 
The overall structure of the ML model for strategy 1 is illustrated in Fig.~\ref{fig:model1}. 
Similar to the 3-label identification in step 1, the PMT waveform features from the prompt trigger are fed into a model based on PointNet++. 
For the neutron-capture vertices, given that the number of neutrons is usually less than 100 for an atmospheric neutrino event, the size of the point-cloud data is fixed to 128 points, with zero-padding if the number of selected neutron-capture candidates is less than 128. 
This point-cloud data of neutrons is fed into an individual Dynamic Graph CNN (DGCNN)-based model ~\cite{DGCNN} . 
Compared to traditional GCNN models, DGCNN is able to recover the neighborhood topology of point clouds with edge information, making it more suitable for describing spatial relationships of sparse point clouds. 
The DCGNN-based model output is then concatenated with the PointNet++-based model output with a set of fully-connected layers.

\subsubsection{Strategy 2}

Instead of reconstructing individual neutron-capture vertices, this strategy merges multiple delayed triggers into one by summing up the charges from multiple selected delayed triggers for each PMT, and taking the FHT value as the earliest among them (Fig.~\ref{fig:strategy2_merge}).  
The two features (FHT and charge) are then used as extra features to the ML model together with features from the prompt trigger. 
This strategy does not rely on neutron-capture reconstruction algorithms and provides a fast and coherent way of utilizing features from both the prompt trigger and the delayed trigger. 
A DeepSphere-based model with the same architecture as used in the 3-label identification is used for the demonstration of this strategy. 
%For both strategies, the final output is a 2-label classifier, with 2 scores representing the probability of an event being $\nu$-like or $\bar{\nu}$-like respectively.
%The models are then trained separately for the $\nu_\mu$ vs $\bar{\nu}_\mu$ and $\nu_e$ vs $\bar{\nu}_e$ identification.   

\begin{figure*}[htbp]
\centering
\includegraphics[width=0.45\textwidth]{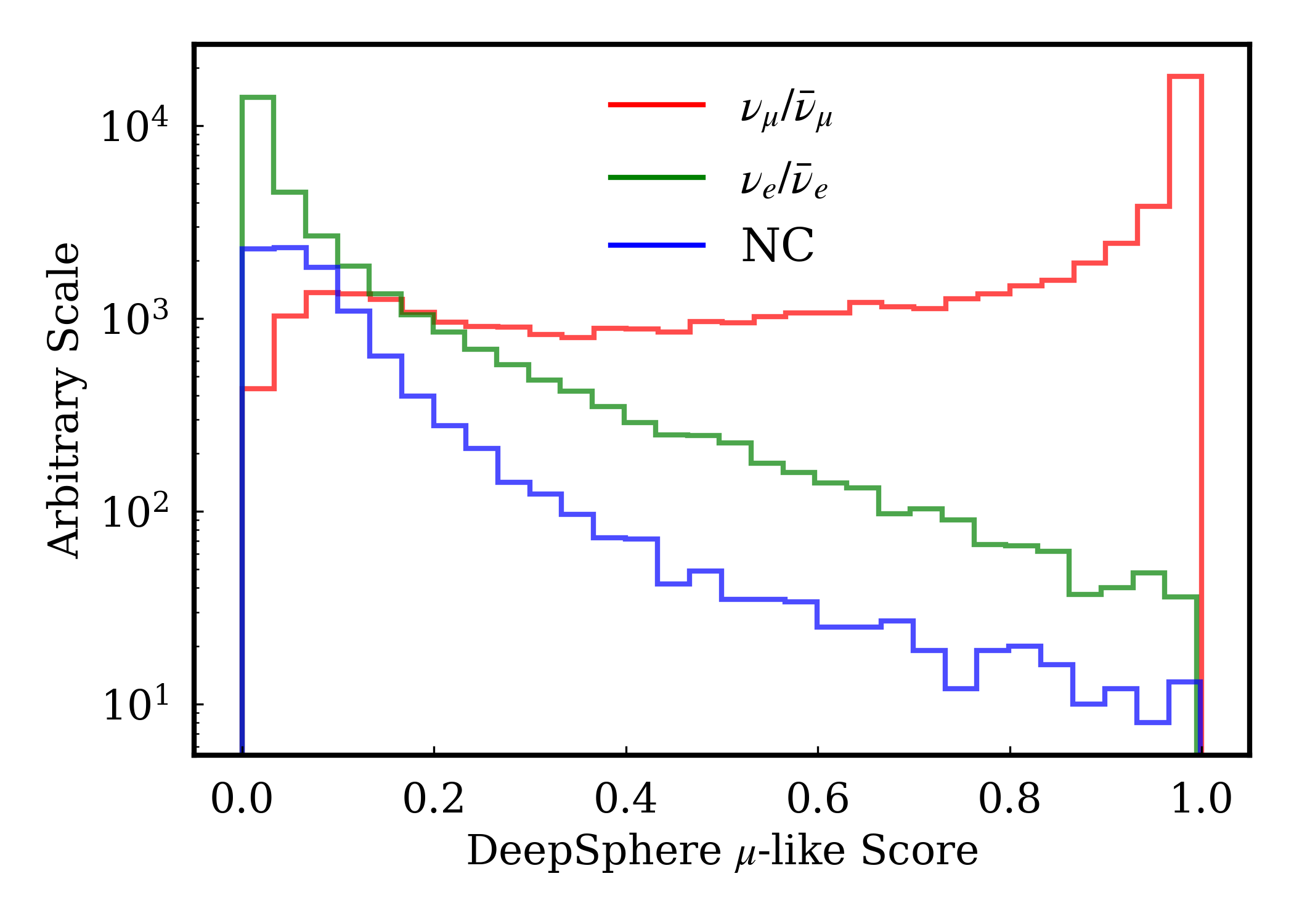}
\includegraphics[width=0.45\textwidth]{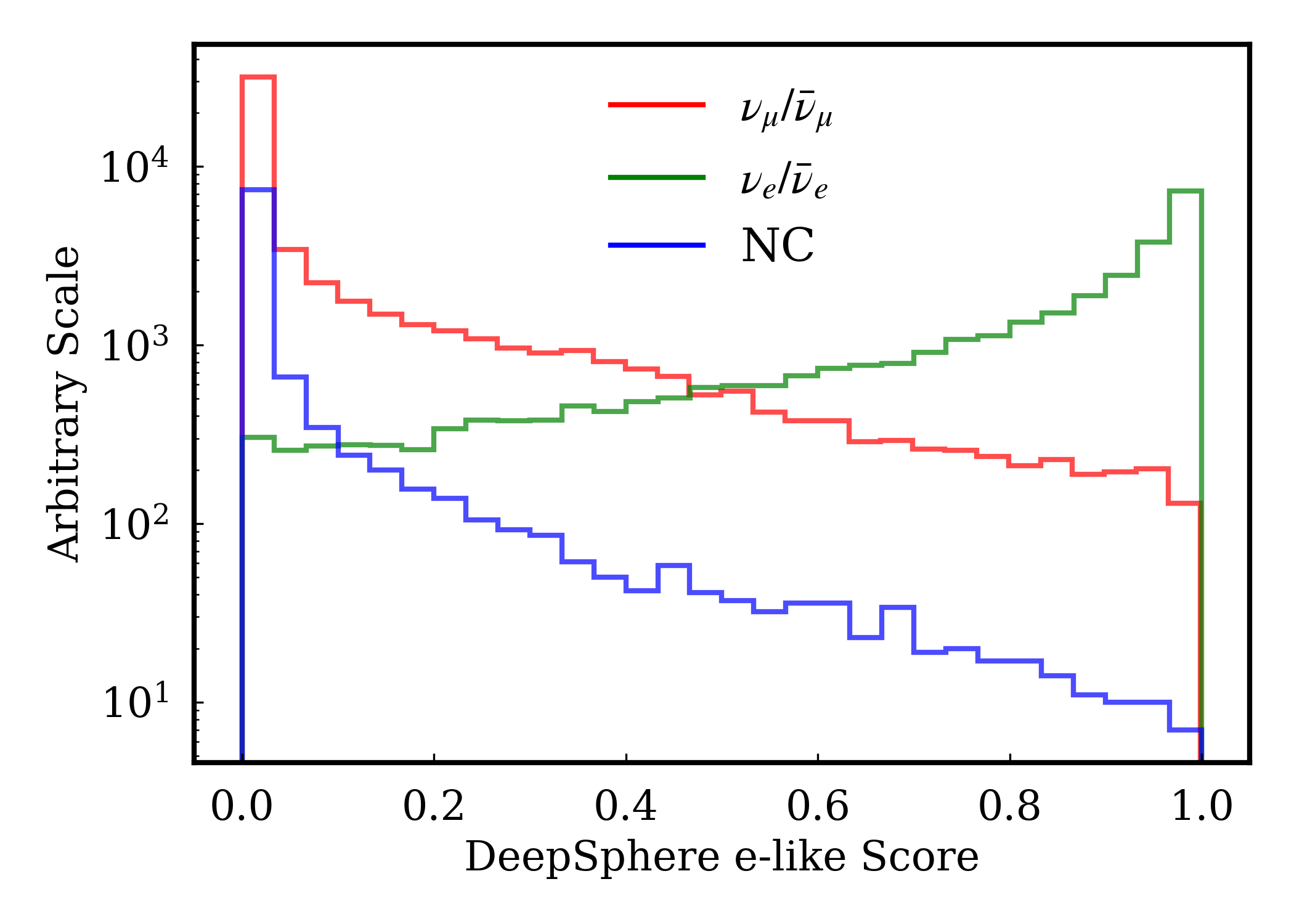}
\includegraphics[width=0.45\textwidth]{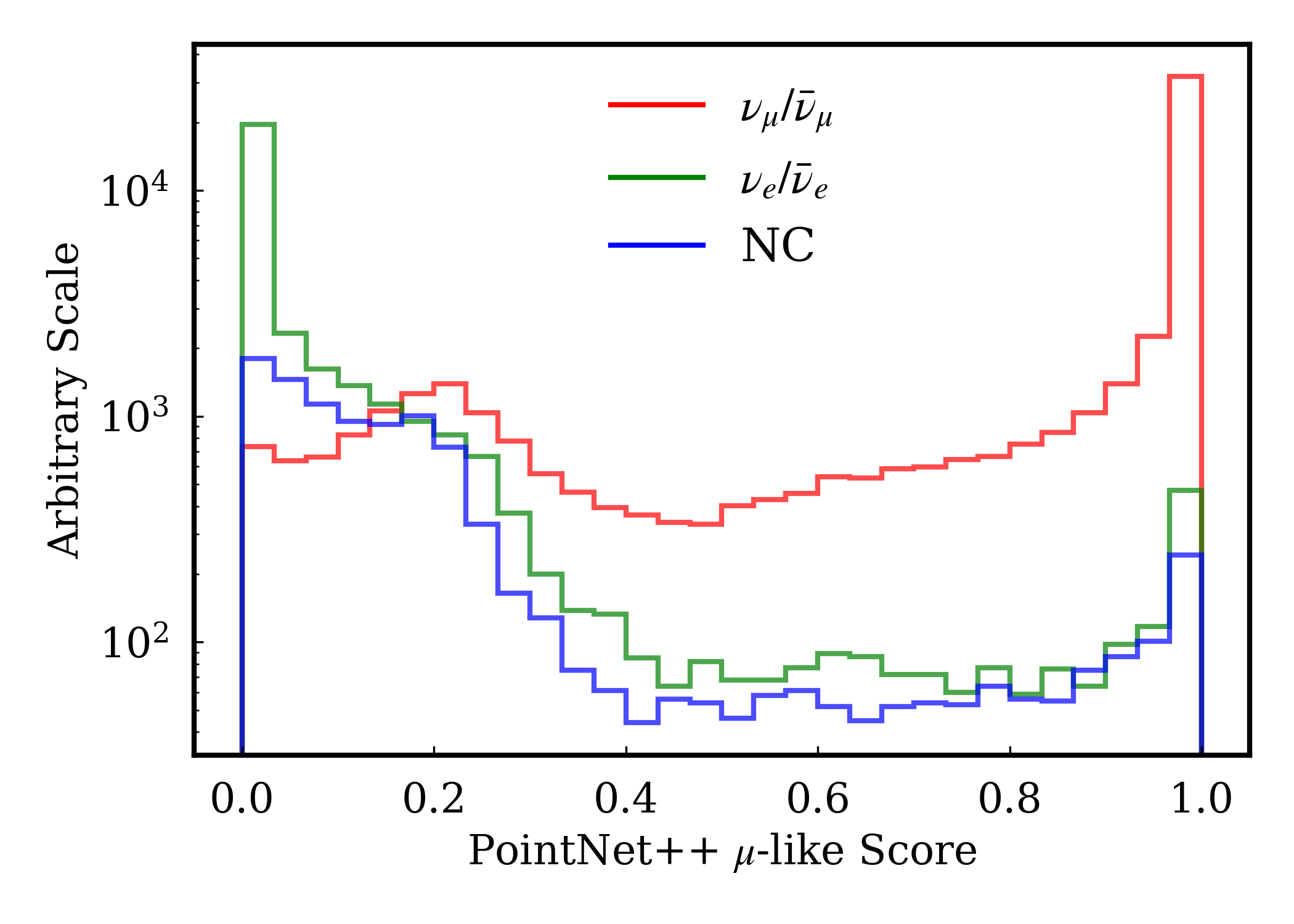}
\includegraphics[width=0.45\textwidth]{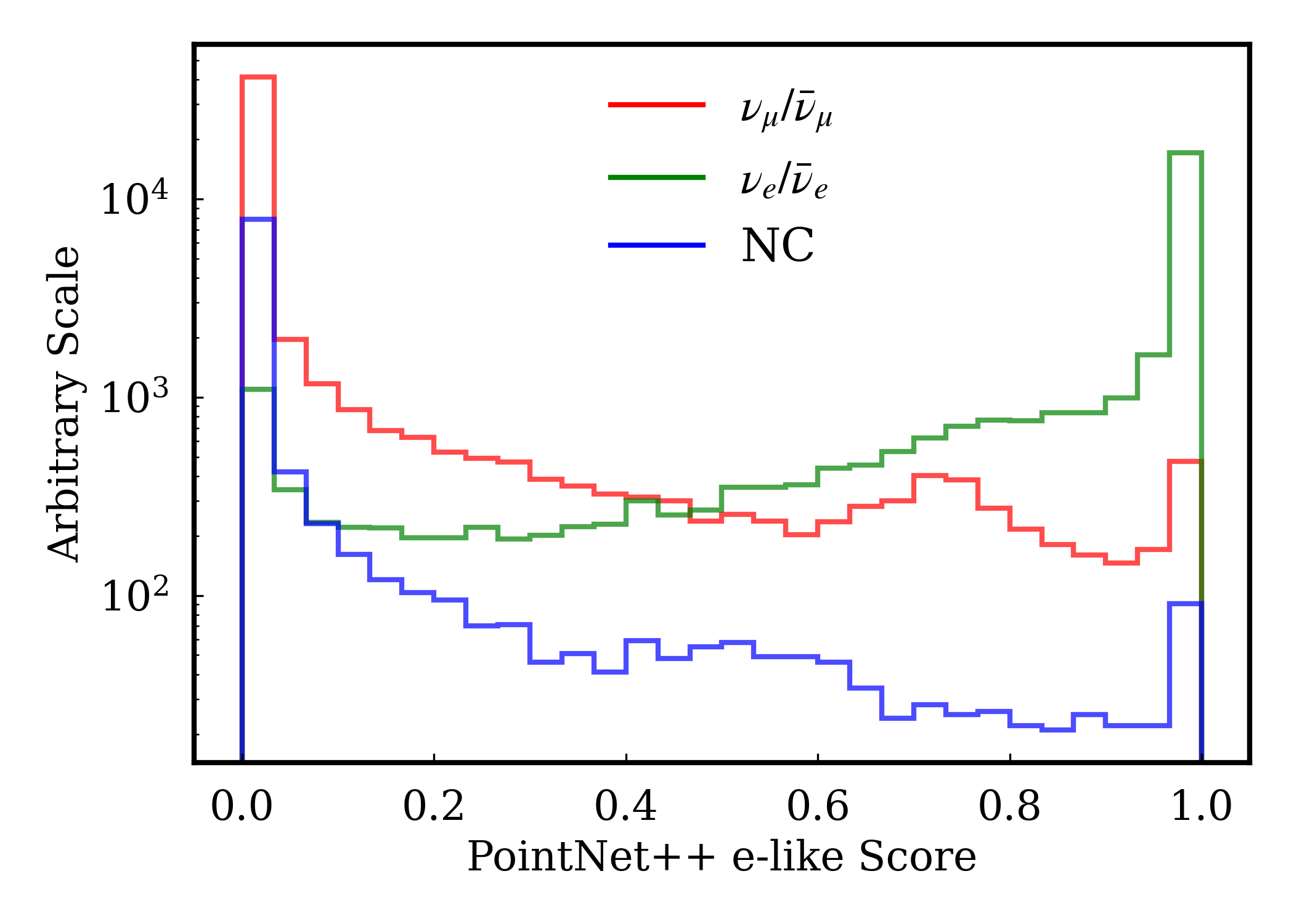}
\caption{Distributions of the $\mu$-like score (left) and $e$-like score (right) by the Deepsphere (top) and PointNet++ (bottom) models. True $\nu_\mu$/$\bar{\nu}_\mu$-CC, $\nu_e$/$\bar{\nu}_e$-CC and NC events are shown in different colors. } 
\label{fig:score_3label}
\end{figure*}

\begin{figure*}[htbp]
\centering
\includegraphics[width=0.45\textwidth]{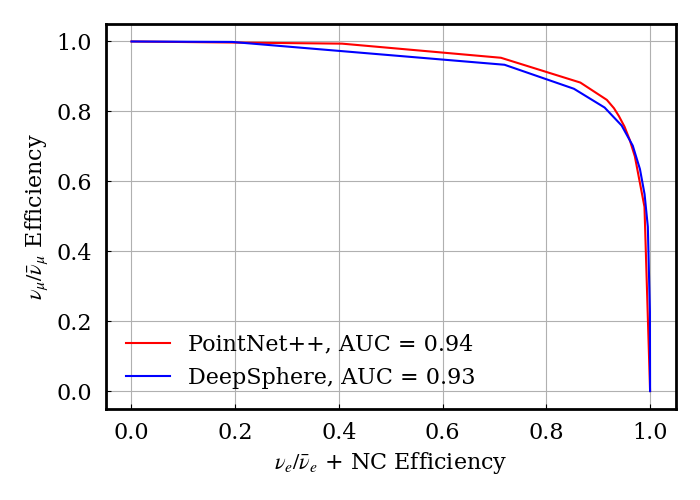}
\includegraphics[width=0.45\textwidth]{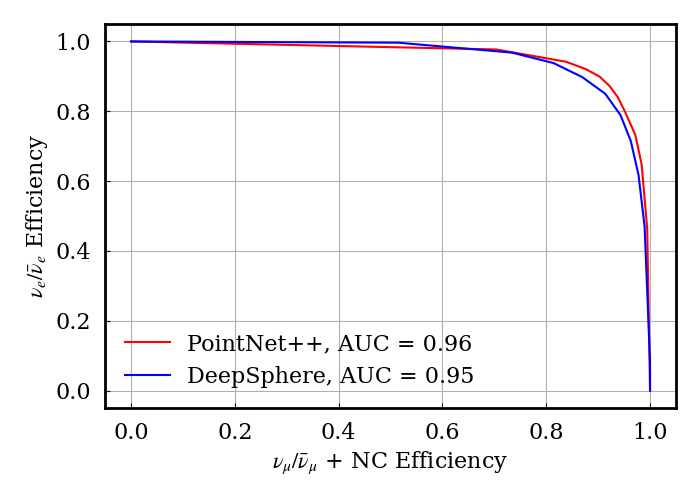}
\caption{ROC curves of the 3-label identification by the two ML models using events across all energies. 
For each ROC plot, the vertical axis shows the selection efficiency of one of the three classes, and the horizontal axis shows the efficiency of the other two combined. 
The corresponding AUC values are also listed. } 
\label{fig:ROC_3label}
\end{figure*}

\section{Performance \label{sec:performance}}

The 3-label and 2-label ML models are trained with the flat sample labeled by event categories.
The trained models are then applied to the Honda flux sample for performance evaluation in a more realistic situation. 
The efficiency, purity, and accuracy of event selection depend on the cuts on the output scores of ML models, which should be optimized by future analyses. 
To evaluate the model performances independent from the choices of selection cuts, the receiver operating characteristic (ROC) curve and the area under the ROC curve (AUC) are used~\cite{ROC_AUC}.  
They have the advantage of independence from the score cut choice, and do not suffer from a class imbalance in the dataset if there is one. 
While the typical ROC curve and AUC are designed for binary classifications, the concepts are extended to an $N$-class classification in this work. 
For a given class $i$, the ROC curve and corresponding AUC score are obtained by converting the $N$-class classification to a ``class $i$ vs others'' binary classification.
The efficiency of class $i$ is defined as the fraction of events in class $i$ with the score associated with the class greater than a certain cut value.
In contrast, the selection efficiency of ``others'' is calculated as the fraction of events with the score lower than the cut value.  
By varying the cut value from 0 to 1, one can calculate the corresponding two efficiencies above and subsequently draw the ROC curve of ``class $i$ vs others'', from which an AUC value is obtained. 
Finally, the total AUC score is defined as the arithmetic mean of all ``class $i$ vs others'' binary classification scores. 
The performance of the 3-label and 2-label models are presented in the following subsections separately. 

\begin{figure}[htbp]
\centering
\includegraphics[width=0.45\textwidth]{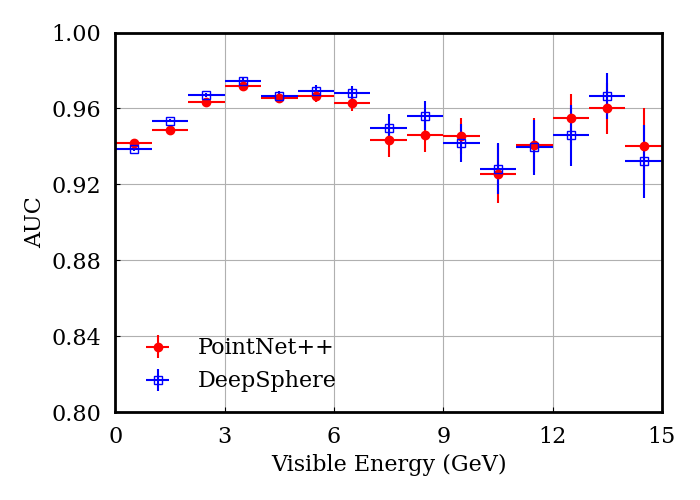}
\caption{The total AUC values of the 3-label identification as functions of visible energy for the two models. Error bars show the statistical uncertainties in each energy bin obtained with the bootstrap method~\cite{bootstrap}.} 
\label{fig:AUC_3label}
\end{figure}

\begin{figure*}[htbp]
\centering
\includegraphics[width=0.45\textwidth]{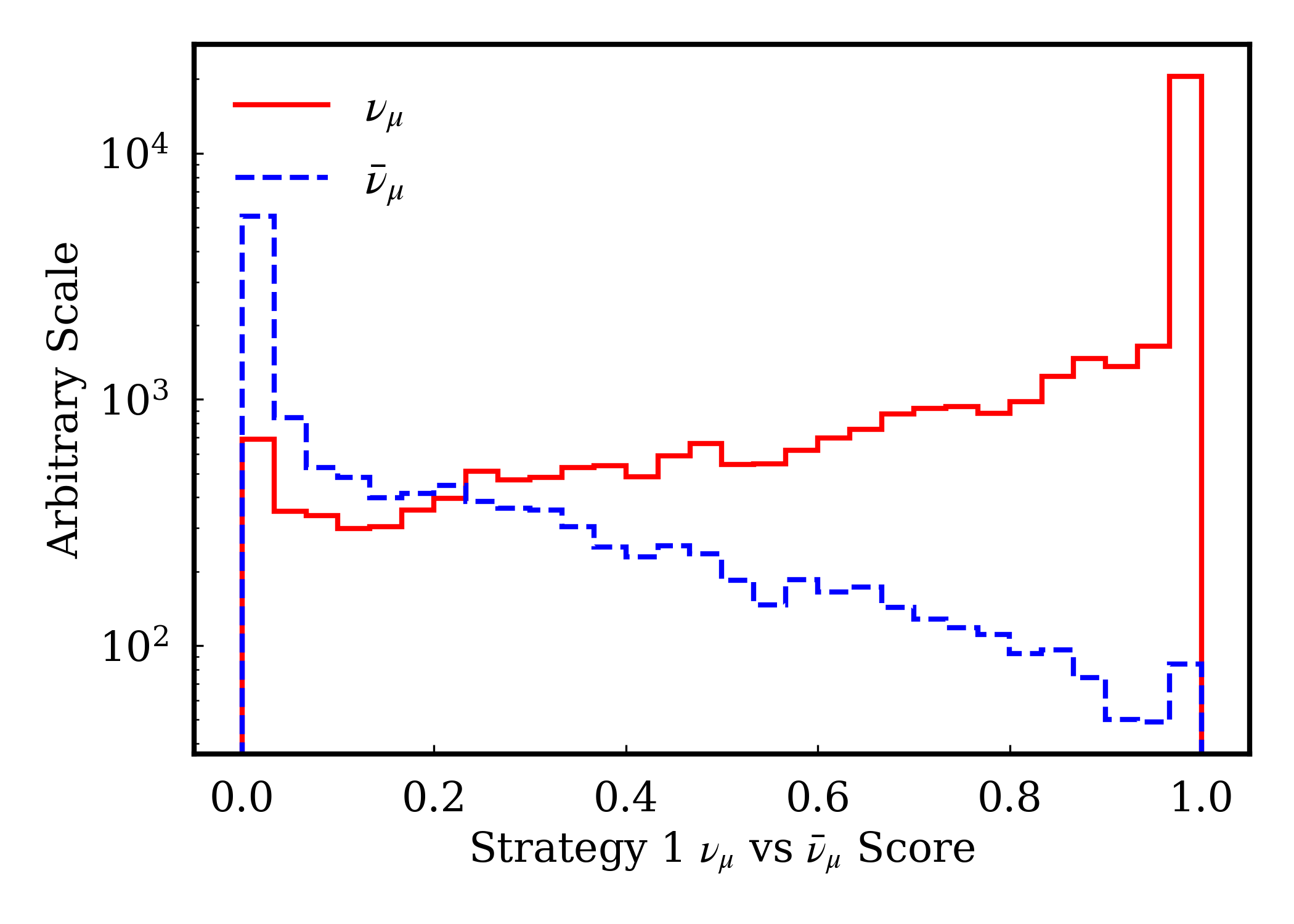}
\includegraphics[width=0.45\textwidth]{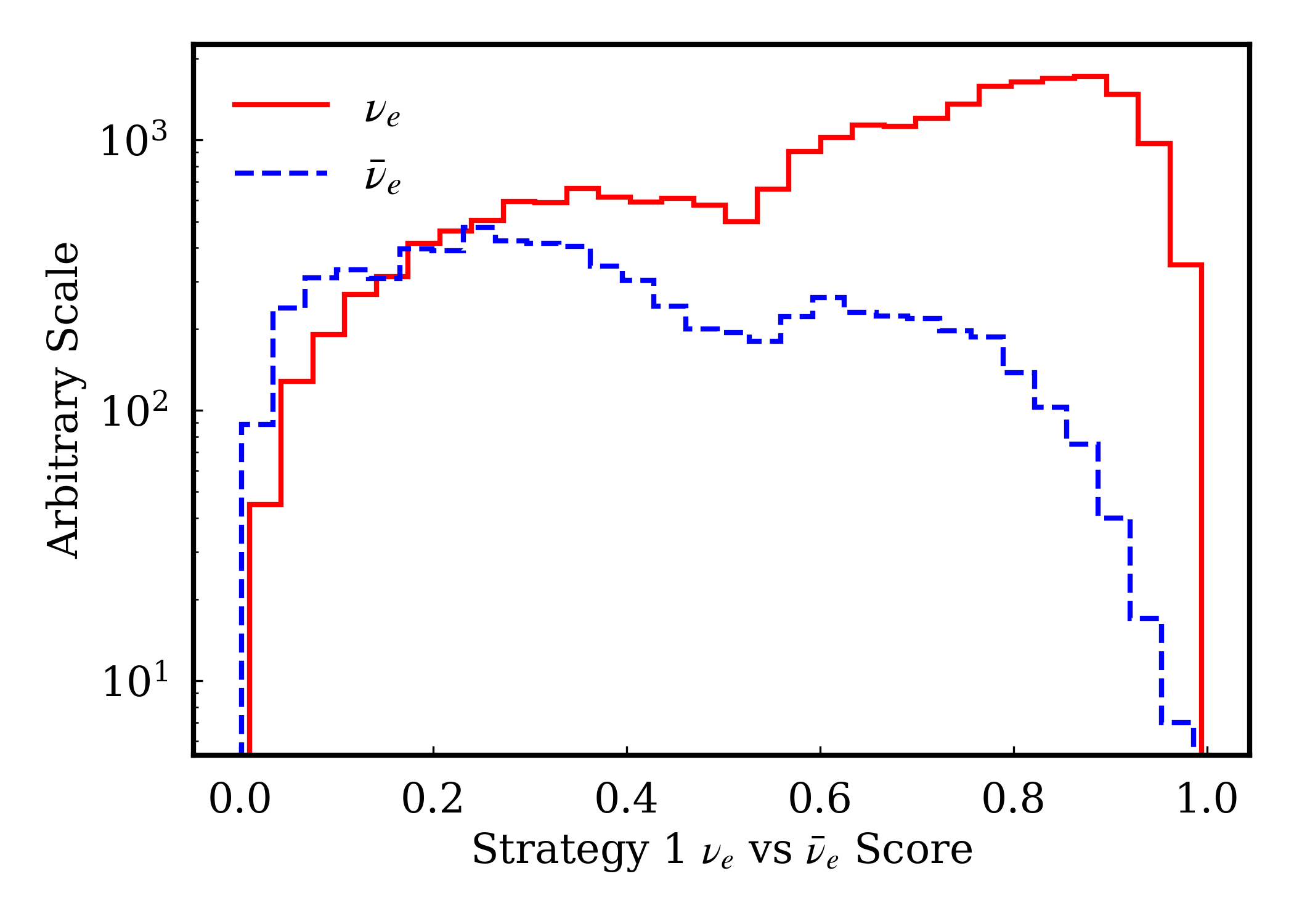}\\
\includegraphics[width=0.45\textwidth]{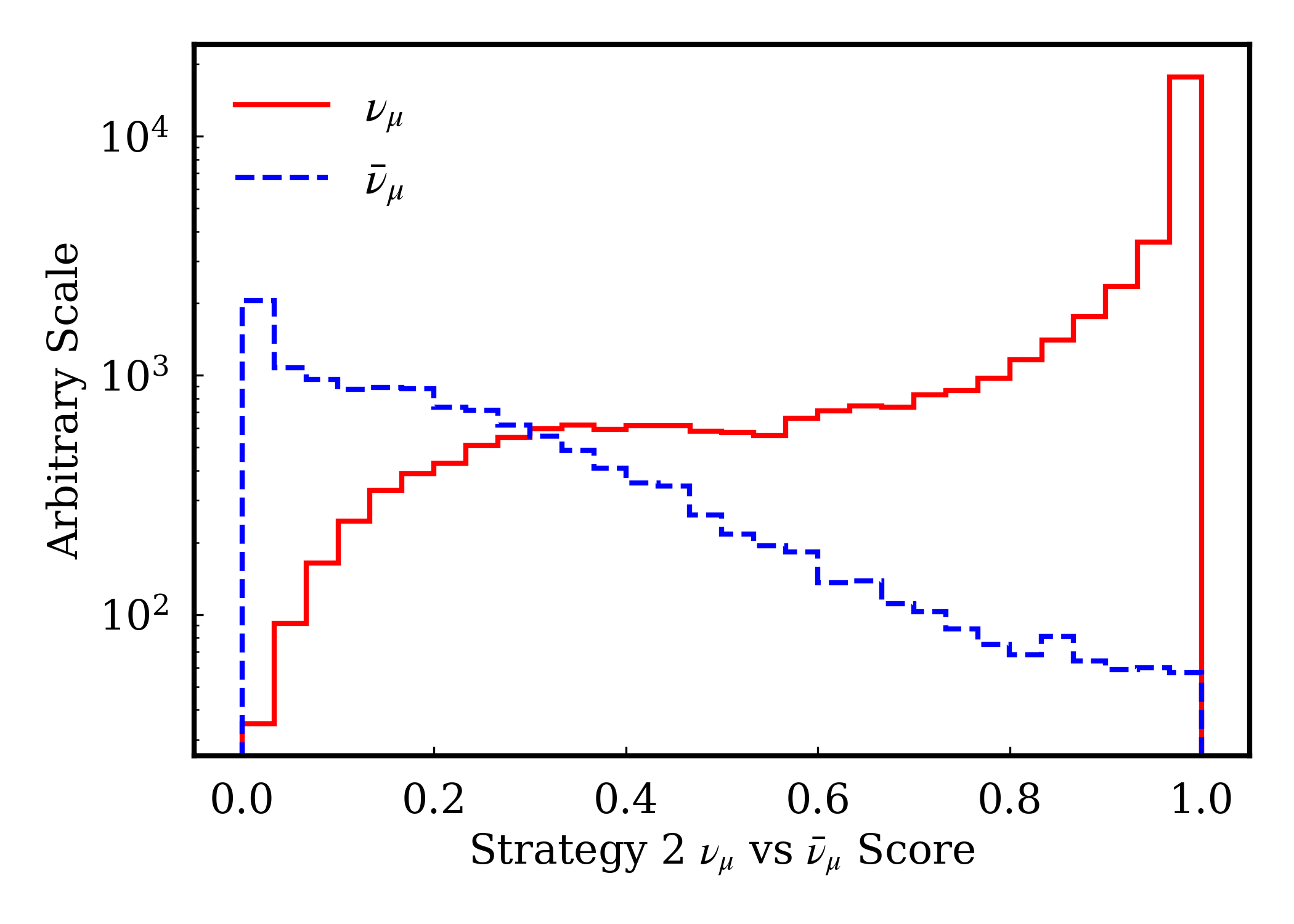}
\includegraphics[width=0.45\textwidth]{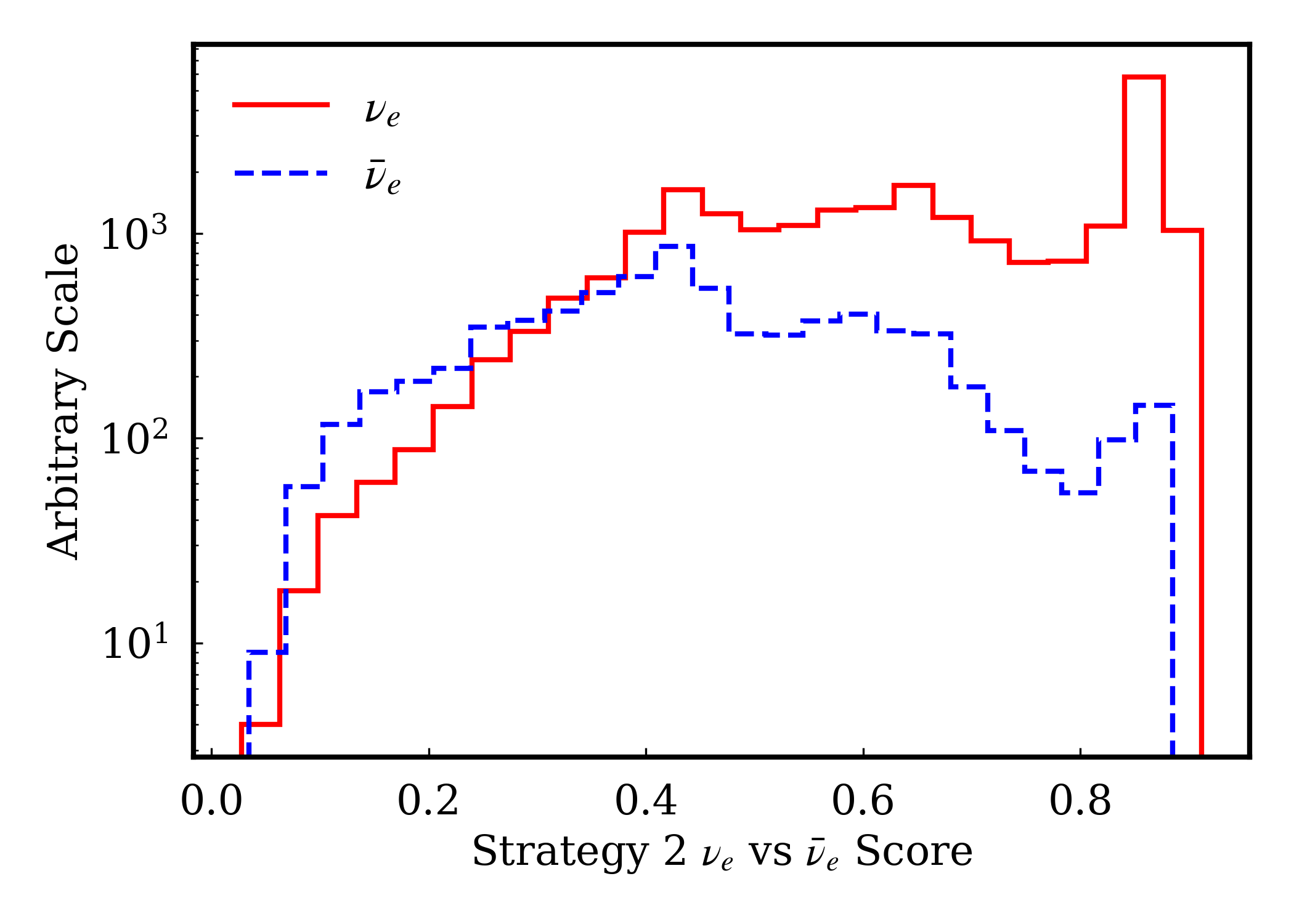}
\caption{\label{fig:nu_vs_antinu} The distributions of $\nu_\mu$ vs $\bar{\nu}_\mu$ (left) score and $\nu_e$ vs $\bar{\nu}_e$ (right) score by strategy 1 (top) and 2 (bottom).}
\end{figure*}

\begin{figure*}[htbp]
\centering
\includegraphics[width=0.45\textwidth]{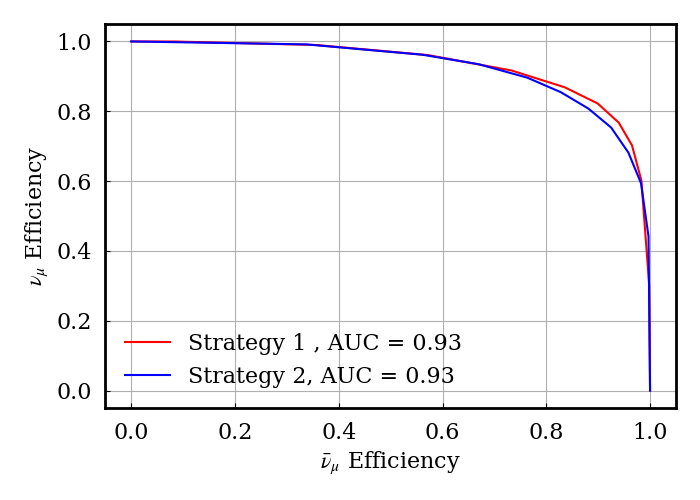}
\includegraphics[width=0.45\textwidth]{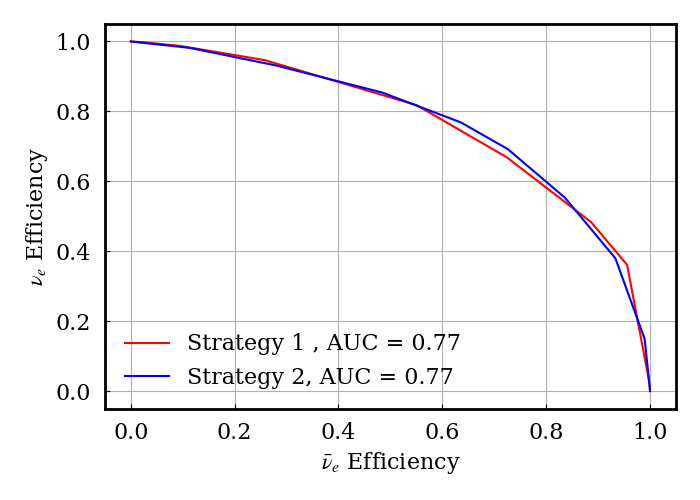}
\caption{ROC curves of the $\nu_\mu$ vs $\bar{\nu}_\mu$ (left) and $\nu_e$ vs $\bar{\nu}_e$ (right) identification.  The vertical axis represents the selection efficiency of neutrinos by varying cut values on the score shown in Fig.~\ref{fig:nu_vs_antinu}, and the horizontal axis represents the efficiency of the antineutrinos. The corresponding AUC values are also listed.} 
\label{fig:ROC_2label}
\end{figure*}

\begin{figure*}[htbp]
\centering
\subfigure[$\nu_\mu$/$\bar{\nu}_\mu$ identification]{
\includegraphics[width=0.45\textwidth]{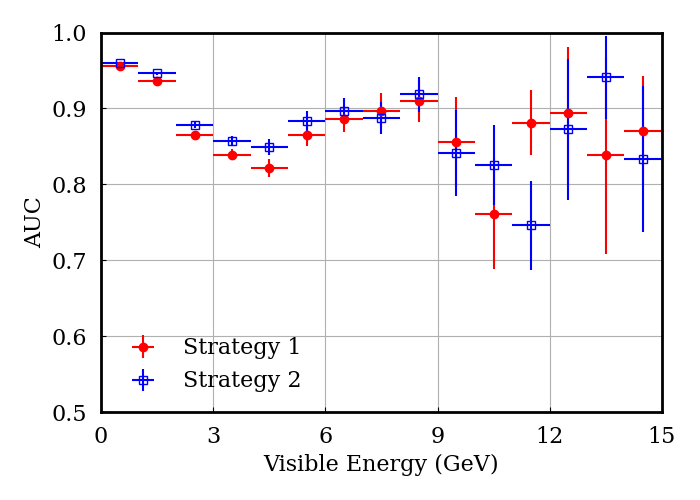}
}
\subfigure[$\nu_e$/$\bar{\nu}_e$ identification]{
\includegraphics[width=0.45\textwidth]{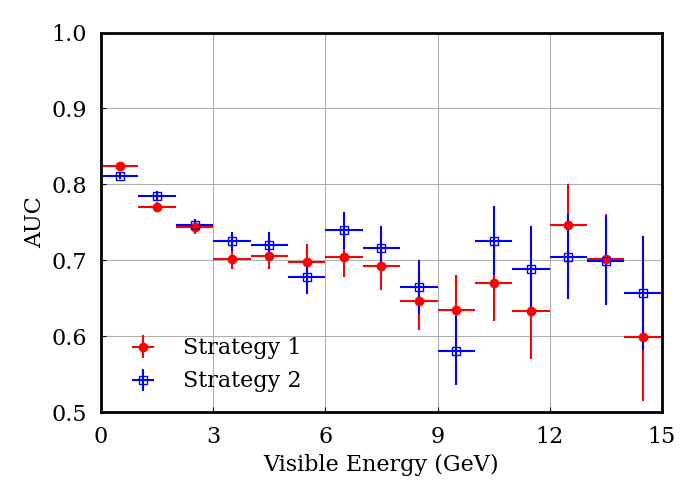}
}
\caption{Comparison of the AUC values for the 2-label (a) $\nu_\mu$/$\bar{\nu}_\mu$ and (b) $\nu_e$/$\bar{\nu}_e$ identifications as functions of visible energy between the two strategies.}
\label{fig:AUC_2label}
\end{figure*}

\subsection{3-label flavor identification}

Fig.~\ref{fig:score_3label} shows the distributions of the $\mu$-like score and $e$-like score output by the two 3-label ML models. 
%(The NC-like scores are not shown since NC is a background.)
(The NC-like scores are not shown since NC is a background to the oscillation analysis and the NC-like scores can be deduced from the $\mu$-like and $e$-like scores.)
Clear separations are observed for the three event classes. 
ROC curves can be obtained from the score distributions, using the ``class $i$ vs others'' method. 
Fig.~\ref{fig:ROC_3label} shows the ROC curves and AUC values for the two ML models using events across all energies.
Such ROC curves are also calculated for events in different visible energy bins and the corresponding total AUC values are obtained by averaging over the AUC value of each class. 
Fig.~\ref{fig:AUC_3label} compares the total AUC values as a function of visible energy for the two ML models.  
The AUC values are rather consistent, indicating that the performances of the two models are very close. 

\subsection{2-label $\nu$/$\bar{\nu}$ identification}

Similar to the 3-label identification, the performances of the $\nu$/$\bar{\nu}$ identification are evaluated in terms of ROC curve and AUC values using the Honda-flux sample. 
The ML model output score distributions are shown in Fig.~\ref{fig:nu_vs_antinu}, where distinct shapes are obtained for neutrinos and antineutrinos. 
The selection efficiencies of (anti)neutrinos for the ROC curve are obtained by selecting events with scores above (below) various cut values. 
The ROC and AUC comparison of the two strategies are shown in Fig.~\ref{fig:ROC_2label} and \ref{fig:AUC_2label} .
The two strategies give similar $\nu$/$\bar{\nu}$ identification performances. 
The remaining differences could be caused by the differences in the models and how the information of the captured neutron candidates is handled.
In addition, the performance for $\nu_\mu$/$\bar{\nu}_\mu$ identification is better than that of $\nu_e$/$\bar{\nu}_e$. 
This could be mainly attributed to more (less) distinctive detector response of muons (electrons) with respect to hadrons in LS. 

\section{Discussion} \label{sec:disscusion}

\subsection{Dependence on event generators}

\begin{figure*}[htbp]
\centering
\subfigure[3-label flavor identiciation]{
\includegraphics[width=0.45\textwidth]{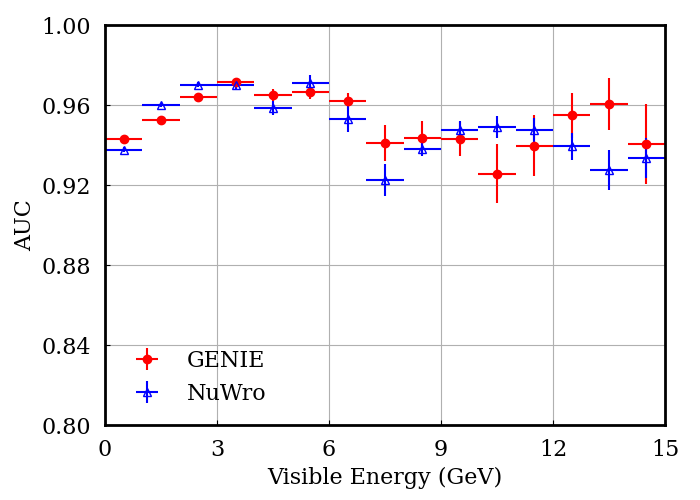}}\\
\subfigure[2-label $\nu_\mu$/$\bar{\nu}_\mu$ identification]{
\includegraphics[width=0.45\textwidth]{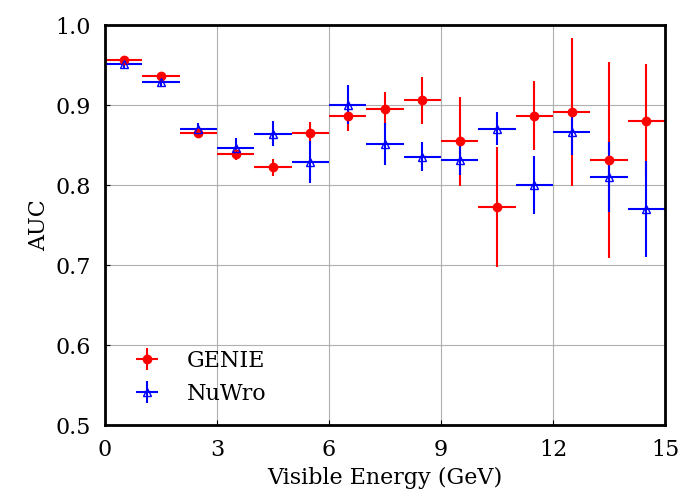}}
\subfigure[2-label $\nu_e$/$\bar{\nu}_e$ identification]{
\includegraphics[width=0.45\textwidth]{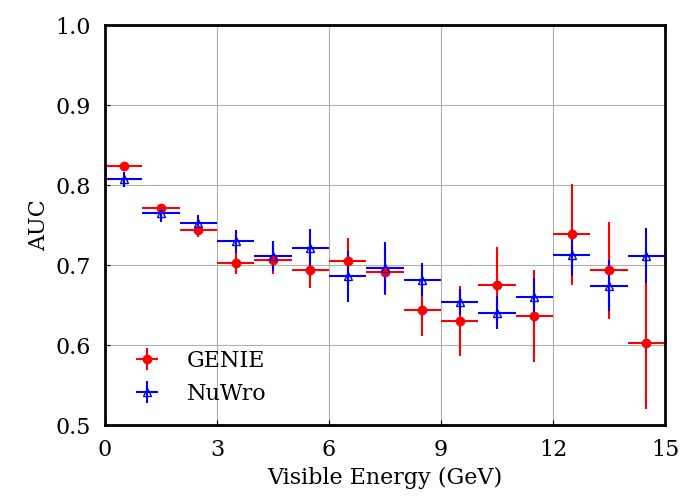}}
\caption{Comparison of the PID performance using the GENIE-generated Honda-flux sample and the NuWro sample. Both (a) the total AUC values for the 3-label flavor identification and the AUC values for the 2-label (b) $\nu_\mu$/$\bar{\nu}_\mu$, (c) $\nu_e$/$\bar{\nu}_e$ identifications are shown. The PointNet++ model and strategy 1 of neutron-capture information handling are used.}
\label{fig:NuWroVsGenie}
\end{figure*}

In this work, event identification utilizes the event topology information reflected by the PMT features in the prompt trigger and the neutron-capture information from the delayed triggers. 
Both pieces of information rely on the simulation of the neutrino-nucleus interaction process, which could be different for various event generators.  
To check the dependence of the PID performance on the neutrino-nucleus interaction simulation, an independent sample was produced using the NuWro event generator, as mentioned in~\ref{sec:det_sim}. 
The models trained with the flat sample produced by GENIE are applied to this NuWro sample.  
Fig.~\ref{fig:NuWroVsGenie} compares the PID performance obtained from the GENIE and NuWro Honda-flux testing samples, using the PointNet++ model and strategy 1 of neutron-capture information handling as an example. 
Despite some small discrepancies, the performance is mostly consistent between the two samples.Similar results were obtained using the Deepsphere model and strategy 2.
This additional check suggests that the ML models in this work do not suffer from apparent dependence on the event generator. 

\begin{figure*}[htbp]
\centering
\includegraphics[width=0.45\textwidth]{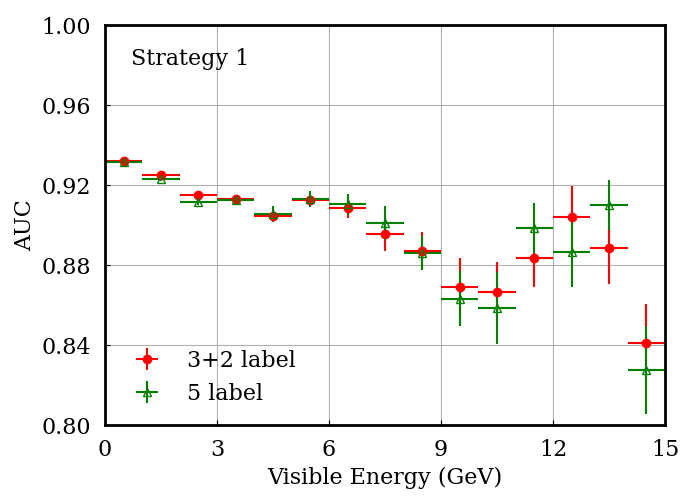}
\includegraphics[width=0.45\textwidth]{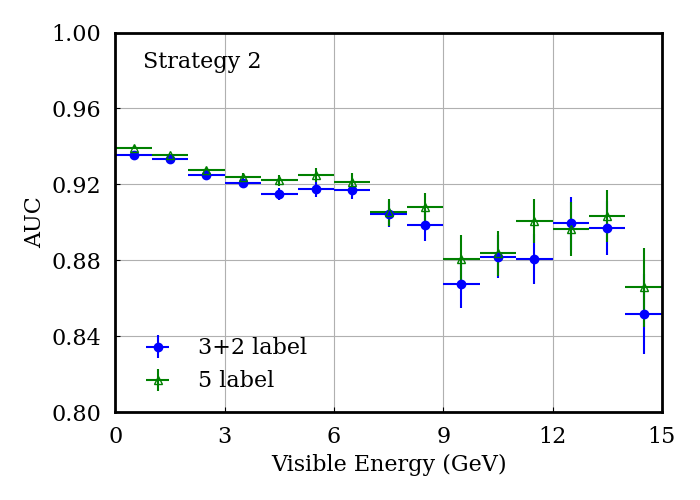}
\caption{Comparison of the PID performance (the total AUC values as functions of visible energy) between 5-label and 3+2-label strategies for the two different neutron information handling strategies. }
\label{fig:5Vs3plus2}
\end{figure*}

\subsection{Selection criteria optimization}
After the flavor and $\nu$/$\Bar{\nu}$ identification, the candidate atmospheric neutrino events shall be categorized into $\nu_{\mu}$-CC, $\bar{\nu}_{\mu}$-CC, $\nu_{e}$-CC, $\bar{\nu}_{e}$-CC and NC categories, while only the former four CC categories are used for the oscillation analysis. 
ML models categorize the event into the class with the largest probability score value by default. 
However, as mentioned previously, this default classification may not be optimal for oscillation analysis.   
Different selections based on the probability scores can be applied, causing events to migrate among the categories, leading to different signal efficiency and background contamination in each category.
For future atmospheric neutrino oscillation analysis, the selection criteria should be optimized to achieve the best sensitivity.
It is for the same reason that the ROC curve and AUC values are chosen to represent the identification performances in this work instead of more conventional benchmarks such as signal efficiency and purity.

\subsection{Alternative classification strategy}
As described in Sec.~\ref{sec:strategy}, the baseline strategy of this work 
is to do the $\mu$-like vs $e$-like vs NC-like 3-label classification first, followed by the $\nu$/$\bar{\nu}$ 2-label classification for the identified $\mu$-like and $e$-like events. 
Thus, this strategy is referred to as the 3+2-label strategy. 
Alternatively, the task can also be achieved by doing the $\nu_e$-CC/$\Bar{\nu}_e$-CC/$\nu_\mu$-CC/$\Bar{\nu}_\mu$-CC/NC  classification all at once using both the PMT features from the prompt trigger and the delayed trigger information, referred to as the 5-label strategy.  
To test this 5-label strategy, the models used for the 2-label classifications are retrained with the same primary trigger and neutron-capture information inputs, and the model architectures remain the same except the outputs are changed to five scores representing the probability of the event being $\nu_\mu / \bar{\nu}_\mu/\nu_e / \bar{\nu}_e$/NC-like. 
The performance is then tested using the same Honda-flux sample and compared with the 3+2-label results using the total AUC scores described in section \ref{sec:performance} for the two different neutron-capture information handling strategies individually. 
To have a fair comparison, 
both the PMT features from the primary trigger and the neutron-capture information are also used for the 3-label classification in the 3+2-label strategy. 
The total average AUC comparisons of the two strategies are shown in Fig.~\ref{fig:5Vs3plus2}, and consistent performances are observed.

\section{Summary \label{sec:sum} }
A machine-learning-based method for the flavor identification and $\nu/\bar{\nu}$ discrimination of
atmospheric neutrinos in a large homogeneous LS detector is presented in this paper.  
It utilizes both the features extracted from PMT waveforms in the prompt trigger and the neutron-capture information in the delayed triggers, taking advantage of LS detectors' low threshold measurements of both the charged lepton and hadrons in a neutrino-nucleus interaction and the capability of high-efficiency neutron tagging.  
PMT features from the prompt trigger can capture different topologies of $e$-like/$\mu$-like/NC-like events, leading to excellent neutrino flavor identification. 
Moreover, they also contain the $y_{vis}$ information, which, combined with the neutron-capture information from the delayed triggers, 
exhibit good separation power between $\nu$ and $\bar{\nu}$.
Different ML models, neutron-capture information handling strategies, and event generators are tested and show consistent performances. 
Using JUNO as an example, we demonstrate for the first time that large homogeneous LS detectors can 
perform identification of the atmospheric neutrino flavor and neutrino/antineutrino discrimination
with their unique advantages. 
Combined with excellent energy resolution and directional measurement capability for neutrinos reported in \cite{Yang:2023rbg}, LS detectors such as JUNO can provide competitive results for future atmospheric neutrino oscillation measurements. 

\section*{Acknowledgements}
This work was partially supported 
by CAS Project for Young Scientists in Basic Research (Grant No.YSBR-099), 
by the National Natural Science Foundation of China (Grant No.12275160, No.12105158, No.12025502, No.12405230),  
%by the Strategic Priority Research Program of the Chinese Academy of Sciences (XDA10010100), 
by National Key R\&D Program of China (Grant No.2024YFE0110500), 
and by Shandong Provincial Natural Science Foundation (Grant No.ZR2022MA062, No.ZR2024QA127). 
%and by the China Postdoctoral Science Foundation (Project No. 2202M713153).  
We would also like to thank the Computing Center of the Institute of High Energy Physics,  Chinese Academy of Science, 
and the Key Laboratory of Particle Physics and Particle Irradiation of Ministry of Education, Shandong University for providing the GPU resources.

\bibliography{main}

\end{document}